\documentclass[11pt]{article}

\usepackage{amssymb}
\usepackage{amstext}
\usepackage{amsfonts}
\usepackage{amsmath}

\newtheorem{theorem}{Theorem}[section]    
\newtheorem{definition}{Definition} 
\newtheorem{claim}[theorem]{Claim}    
\newtheorem{lemma}[theorem]{Lemma}    
\newcommand{\qed}{\hfill{$\rule{6pt}{6pt}$}} 
\newenvironment{proof}{\noindent{\bf Proof}:}{\qed}

\newcommand{\defeq}{\stackrel{\Delta}{=}}

\newcommand{\E}{{\mathbb{E}}}

\newcommand{\C}{{\bf{C}}}

\newcommand{\Good}{{\mathsf{GOOD}}}

\newcommand{\suppress}[1]{}
\newcommand{\tx}{{\tilde{X}}}
\newcommand{\teps}{{\tilde{\epsilon}}}
\newcommand{\cA}{{\mathcal{A}}}

\title{\bf Towards a classical proof of exponential lower bound for $2$-probe smooth codes}
\author{ 
Rahul Jain \\ 
U.\ Waterloo \thanks {Computer Science Division,
University of Waterloo, 200 University Ave.\ W., Waterloo, ON N2L 3G1,
Canada. Email: {\sf rjain@cs.uwaterloo.ca}. This work was partly 
done while the author was at University of California, Berkeley,
CA~94720, USA where it was supported by an Army Research Office
(ARO), North California, grant number DAAD 19-03-1-00082. 
}
}

\date{}

\begin{document}

\maketitle

\begin{abstract}
Let $\C: \{0,1\}^n \mapsto \{0,1\}^m$ be a code encoding an $n$-bit 
string into an $m$-bit string. Such a code is called a $(q, c,
\epsilon)$ {\em smooth code} if there exists a decoding algorithm which
while decoding any bit of the input, makes at most $q$ probes on the
code word and the probability that it looks at any location is at most
$c/m$. The error made by the decoding algorithm is at most
$\epsilon$. Smooth codes were introduced by Katz and
Trevisan~\cite{luca:ldc} in connection with {\em Locally decodable
codes}.

For $2$-probe smooth codes Kerenedis and de Wolf~\cite{ronald:ldc} have shown exponential
in $n$ lower bound on $m$ in case $c$ and $\epsilon$ are
constants. Their lower bound proof went through quantum arguments and
interestingly there is no completely classical argument as yet for the
same (albeit completely classical !) statement.

We do not match the bounds shown by Kerenedis and de Wolf but however
show the following. Let $\C: \{0,1\}^n \mapsto \{0,1\}^m$ be a $(2,c,
\epsilon)$ smooth code and if $\epsilon \leq \frac{c^2}{8n^2}$, then 
$m \geq 2^{\frac{n}{320c^2} - 1}$. We hope that the arguments and
techniques used in this paper extend (or are helpful in making similar
other arguments), to match the bounds shown using quantum arguments.
More so, hopefully they extend to show bounds for codes with greater
number of probes where quantum arguments unfortunately do not yield
good bounds (even for $3$-probe codes).
\end{abstract}

\section{Introduction}
We define smooth codes in a very similar way as defined by Katz and
Trevisan~\cite{luca:ldc} as follows:

\begin{definition} 
\label{def:smooth}
Let $c > 1, 1/2 > \epsilon \geq 0$ and $q$ be an
integer. We call a code $\C: \{0,1\}^n \mapsto \{0,1\}^m$ to be a $(q,
c, \epsilon)$ smooth code if there exists a non-adaptive probabilistic decoding algorithm $\cA$ such that:

\begin{enumerate}
\item For every $x \in \{0,1\}^n$ and every $i \in [n]$, we have:
$$ \Pr [\cA(\C(x),i) = x_i] \geq 1 - \epsilon$$ 
\item For every $i \in [n]$ and every $j \in [m]$, we have,
$$ \Pr [\cA(.,i) \mbox{'reads index'} j] \leq c/m $$
\item In every invocation $\cA$ reads at most $q$ indices of the code non-adaptively.

\end{enumerate}

\end{definition}
{\bf Remark: } Here we are considering only non-adaptive codes,
however for constant probe codes, bounds for non-adaptive codes also
imply bounds (up to constants) for adaptive codes~\cite{luca:ldc}.
  
Katz and Trevisan defined smooth codes in the context of locally
decodable codes (ldc) and showed that existence of ldcs imply
existence of smooth codes. Therefore lower bounds for smooth codes
imply lower bounds for ldcs. However in our case this is not the case
since the error that we are considering is much smaller and we are
letting the smoothness parameter to be constant. Lower bounds for
smooth codes when the error is allowed to be constant also imply
bounds for corresponding ldcs. 

For smooth codes, Kerenedis and de Wolf~\cite{ronald:ldc}, using
interesting quantum information theoretic arguments, showed
exponential in $n$ lower bound on $m$ where they let $c, \epsilon$ to
be constants. This is one of the few nice examples of quantum arguments
leading to classical results. However till now no completely classical
argument for the same result is known. Unfortunately the quantum
arguments of~\cite{ronald:ldc} had a drawback that they could not be
extended to imply similar bounds for smooth codes for higher number of
probes, for instance these arguments fail to lead interesting bounds
even for $3$-probe smooth codes.

We attempt here a completely classical argument for showing
exponential lower bound for $2$-probe smooth codes but we fall short in
terms of showing it for constant error. The result we show for smooth
codes is the following:

\begin{theorem}
\label{thm:main}
Let $\C: \{0,1\}^n \mapsto \{0,1\}^m$ be a $(2, c, \epsilon)$ smooth code and $\epsilon \leq \frac{c^2}{8n^2}$. Then, $m \geq 2^{\frac{n}{320c^2} - 1}$.
\end{theorem}

We hope that although the result here falls short of the desirable,
the arguments presented here could be extended or similar arguments be
made to match, via purely classical arguments, the bounds shown
by~\cite{ronald:ldc} and also more importantly in deriving interesting
bounds for codes with higher number of probes (in particular for
$3$-probe codes).
 
\section{Preliminaries} In this section we briefly review some of the
information theory facts that will be useful for us in our proofs in
the next section. For a good introduction to information theory,
please refer to the fine book by Cover and
Thomas~\cite{cover:infotheory}. We let our random variables to be
finite valued. Let $X, Y$ be random variables. We will let $H(X),
H(X|Y)$ represent the {\em entropy} of $X$ and the {\em conditional
entropy} of $X$ given $Y$. We let  $I(X:Y) \defeq H(X) + H(Y) - H(XY)
= H(X) - H(X |Y)$ represent the {\em mutual information} between $X$ and $Y$.  We
will use the fact $I(X:Y) \geq 0$, alternately $H(X) + H(Y) \geq
H(XY)$, alternately $H(X) \geq H(X | Y ) $, several times without
explicitly mentioning it. We will also use the {\em monotonicity of
entropy} i.e. $H(XY) \geq H(X)$, alternately $H(Y) \geq I(X:Y)$
several times without explicitly mentioning it. Let $X$ be an $m$
valued random variable, then it follows easily that $H(X) \leq \log_2
m$ (below we always take logarithm to the base 2). 

For random variables $X_1, \ldots, X_n$, we have the following {\em
chain rule of entropy}:
\begin{equation} 
\label{eq:chainent}
H(X_1, \ldots X_n) = \sum_{i=1}^n H(X_i | X_1 \ldots X_{i-1})
\end{equation}

Similarly for random variables $X_1, \ldots, X_n, Y$, we have the
following {\em chain rule of mutual information}:
\begin{equation} 
\label{eqn:chain}
I(X_1 \ldots X_n : Y ) = \sum_{i= 1}^n
I(X_i : Y | X_1 \ldots X_{i-1}) 
\end{equation}
Let $X, Y, Z$ be random variables. Then we have the following important
{\em monotonicity relation} of mutual information:

\begin{equation}
\label{eqn:mono}
I(XY: Z ) \geq I(X:Z)
\end{equation}

All the above mentioned relations also hold for conditional random
variables for example, for random variables $X,Y,Z, I(X:Y |Z) \geq 0,
H(XY |Z) \geq H(X|Z)$ and so on. Again we may be using the conditional
versions of the above relations several times without explicitly
mentioning it. 

For correlated random variables $X,
Y$, we have the following Fano's inequality. Let $ \epsilon \defeq \Pr[X
\neq Y]$ and let $|X|$ represent the size of the range of $X$. Then  
\begin{equation}
\label{eqn:fano}
H(\epsilon) + \epsilon \log (|X| -1) \geq H(X | Y)
\end{equation}

For $0 \leq p \leq 1/2$, we have the bound $H(p) \leq 2\sqrt{p}$.

\section{Proof of Theorem~\ref{thm:main}}
Let $X \defeq X_1 \ldots X_n$ be a random variable uniformly
distributed in $\{0,1\}^n$ (corresponding to the input being encoded)
and $X_i$ correspond to the $i$-th bit of $X$. This implies that
$X_i$'s are distributed independently and uniformly in $\{0,1\}$. Let
$Y \defeq Y_1 \ldots Y_m$ be a random variable (correlated with $X$)
corresponding to the code, i.e $Y = \C(X)$. Here $Y_j, j \in [m]$
corresponds to the $j$-th bit of the code.



Let $\cA$ be as in Definition~\ref{def:smooth}.  Let $0 \leq \epsilon
< 1/2$, for $i \in [n]$ let $E^{\epsilon}_i$ be the graph on $[m]$
consisting of edges $(j,k)$ such that,

\begin{equation} \label{eqn:pr} \Pr[\cA(\C(X),i) = X_i | \cA \mbox{
reads } (Y_j, Y_k) ] \geq 1 - \epsilon 
\end{equation} 

Following interesting fact can be shown using arguments of Katz and
Trevisan~\cite{luca:ldc}:

\begin{lemma}
\label{lem:largematching}
Let $\C: \{0,1\}^n \mapsto \{0,1\}^m$ be a $(2,c,\epsilon)$ smooth
code. Let $E^{2\epsilon}_i$ be as described above. Then for each $i \in [n]$,
$E^{2\epsilon}_i$ has a matching $M_i$ of size at least $\frac{m}{4c}$.
\end{lemma}

\begin{proof}
Using the definition of smooth code we have,

\begin{eqnarray*}
1 - \epsilon & \leq & \Pr[\cA(\C(X),i) = X_i | \cA(\C(X),i) \mbox{ reads }
E^{2\epsilon}_i ]\Pr[\cA(\C(X),i) \mbox{ reads } E^{2\epsilon}_i] \\ 
& + &  \Pr[\cA(\C(X),i) = X_i | \cA(\C(X),i) \mbox{ reads complement of } E^{2\epsilon}_i
]\Pr[\cA(\C(X),i) \mbox{ reads complement of } E^{2\epsilon}_i] \\
& \leq & \Pr[\cA(\C(X),i) \mbox{ reads } E^{2\epsilon}_i] + ( 1- 2\epsilon)
(1 - \Pr[\cA(\C(X),i) \mbox{ reads } E^{2\epsilon}_i]) 
\end{eqnarray*}
This implies $ \Pr[\cA(\C(X),i) \mbox{ reads }  E^{2\epsilon}_i] \geq  1/2$. 
For an edge $e \in E^{2\epsilon}_i$, let $P_e \defeq \Pr[\cA(\C(X),i)
\mbox{ reads } e]$. This implies $\sum_{e \in E^{2\epsilon}_i} P_e
\geq 1/2$. Furthermore since $\C$ is a $(2,c,\epsilon)$ smooth code,
for every $j \in [m]$,  it implies $\sum_{e \in E^{2\epsilon}_i| j \in
e}P_e \leq c/m$. Let $V$ be a vertex cover of
$E^{2\epsilon}_i$. Therefore,
$$ 1/2 \leq \sum_{e \in E^{2\epsilon}_i| e \cap V \neq \emptyset }
\leq \sum_{j \in V} \sum_{e \in E^{2\epsilon}_i | j \in e }P_e \leq |V|c/m$$
This implies that minimum vertex cover of $E^{2\epsilon}_i$ has size
at least $m/2c$. This now implies that $E^{2\epsilon}_i$ has a
matching of size at least $m/4c$.
\end{proof}

We start with the following claim.

\begin{claim}
\label{claim:inf}
Let $(j,k) \in M_i$ and $\epsilon' \defeq \sqrt{8 \epsilon} $.  Then,
$ I(X_i : Y_j Y_k ) \geq 1 - \epsilon'$.
\end{claim}
\begin{proof}
\begin{eqnarray*}  I(X_i : Y_j Y_k ) & = & H(X_i) - H(X_i | Y_jY_k) \\ 
& \geq & 1 - H(2\epsilon) \mbox{ (from(\ref{eqn:fano}) and (\ref{eqn:pr}))}\\ 
& \geq & 1 - \sqrt{8\epsilon}  \mbox{  (from the bound $H(p) \leq 2 \sqrt{p}$)}
\end{eqnarray*}
\end{proof}

We make the following claim which roughly states that the information
about various $X_i$s do not quite go into the individual bits of
$Y$. For $i \in [n]$ let, $\tx_i \defeq X_1 \ldots X_{i-1}$.

\begin{claim} 
\label{claim:good}
$$ \sum_{i \in [n]} \sum_{(j,k) \in M_i} I(X_i : Y_j | \tx_i) +
I(X_i : Y_k | \tx_i) \leq m $$
\end{claim}\begin{proof}
\begin{eqnarray*}
\sum_{i \in [n]} \sum_{(j,k) \in M_i} I(X_i : Y_j | \tx_i) +
I(X_i : Y_k | \tx_i)   & \leq &  \sum_{i \in [n]} \sum_{j \in [m]}
I(X_i : Y_j | \tx_i) \mbox{ (since $M_i$s are matchings)} \\
& = & \sum_{j \in [m]} \sum_{i \in [n]}   I(X_i : Y_j | \tx_i) \\
&  = &  \sum_{j \in [m]} I(X : Y_j ) \mbox{ (from (\ref{eqn:chain}))}\\ 
& \leq &  m  \mbox{ (since $\forall j \in [m], Y_j$ is a binary random variable)} 
\end{eqnarray*}
\end{proof}

We now have the following claim which roughly states that for a
typical edge $(j,k) \in M_i$ there is a substantial increase in
correlation between $Y_j$ and $Y_k$ after conditioning on $X_i$.
\begin{claim} \label{claim:nice}
Let $\epsilon' \leq \frac{c}{n}$. Then,
$$ \E_{i \in_U [n], (j,k) \in_U M_i} [I(Y_j : Y_k | X_i\tx_i) - I(Y_j :
Y_k|\tx_i) ] \geq 1 - 5c/n $$
\end{claim}
\begin{proof}
Let $(j,k) \in M_i$. Since $X_i$ and $\tx_i$ are independent
random variables, this implies $I(X_i : \tx_i) = 0$ and we get:
\begin{eqnarray*}
I(X_i : Y_jY_k)  & \leq &  I(X_i : \tx_iY_jY_k)  \mbox{ (from (\ref{eqn:mono}))}\\
& = & I(X_i : \tx_i) + I(X_i :Y_jY_k |  \tx_i) \mbox{ (from
(\ref{eqn:chain}))}\\
& = &  I(X_i :Y_jY_k |  \tx_i) \\
& = & I(X_i : Y_j | \tx_i) + I(X_i : Y_k | \tx_i) + I(Y_j : Y_k |
X_i\tx_i) - I(Y_j : Y_k|\tx_i) \mbox{ (from (\ref{eqn:chain}))}\\
\end{eqnarray*}
From Claim~\ref{claim:inf}  we get,
\begin{eqnarray*}
 (1 - \epsilon') \sum_i |M_i|  & \leq &  \sum_i \sum_{(j,k) \in M_i}
I(X_i : Y_jY_k) \\
& \leq &  \sum_i \sum_{(j,k) \in M_i}  I(X_i : Y_j |
\tx_i) + I(X_i : Y_k | \tx_i) + I(Y_j : Y_k | X_i\tx_i) - I(Y_j :
Y_k|\tx_i) 
\end{eqnarray*}
Claim~\ref{claim:good} now implies:
\begin{eqnarray*}
& & \sum_i \sum_{(j,k) \in M_i} I(Y_j : Y_k | X_i\tx_i) - I(Y_j :
Y_k|\tx_i)  \geq  (1 - \epsilon') \sum_i |M_i| - m  \\
& \geq & (\sum_i |M_i|) ( 1- \epsilon' - \frac{m}{\sum_i |M_i|})  \geq
(\sum_i |M_i|)  ( 1 - c/n - 4c/n )  \mbox{ (from Lemma~\ref{lem:largematching})}
\end{eqnarray*}
\end{proof}

Applying Markov's inequality on the above claim we get:

\begin{claim} \label{claim:markov} 
Let $0 < \delta_1, \delta_2 \leq 1$. There exists a set $\Good \subseteq [n]$ and sets
$\Good_i \subseteq M_i$ such that:
\begin{enumerate}
\item $|\Good| \geq (1 - \delta_1) n$ and $i \in \Good, \E_{(j,k) \in M_i} [I(Y_j : Y_k | X_i\tx_i) - I(Y_j :
Y_k|\tx_i) ] \geq 1 - \frac{5c}{\delta_1n} $ 
\item $\forall i \in \Good, |\Good_i| \geq  (1 - \delta_2) |M_i|$ and for $(j,k) \in \Good_i,  I(Y_j : Y_k | X_i\tx_i) - I(Y_j :
Y_k|\tx_i) \geq 1 - \frac{5c}{\delta_1\delta_2n}$ 
\end{enumerate}
\end{claim}

Let $\delta_1 = \delta_2 = 1/2$. Let $\teps \defeq \frac{20c}{n}$. Therefore for $i \in \Good$ and $(j,k) \in \Good_i$ we have from above,
\begin{equation}
\label{eqn:main}
I(Y_j : Y_k | X_i\tx_i) - I(Y_j :
Y_k|\tx_i) \geq  1 - \tilde{\epsilon}
\end{equation}

We
fix $\Good$ to have exactly $\frac{1}{2\teps} - 2 $
elements. For $i \in \Good$, let $a_i$ be the index of $i$ in $\Good$. For $i 
\notin \Good$, let $a_i$ be the index of largest $i' < i$ in $\Good$. For $j \in [m], i \in [n]$, let $S^i_j \defeq \{l \in [m] : H(Y_j | Y_lX_i\tx_i) \leq a_i \teps\}$. Let $S_j^0 \defeq \{j\}$.    

We show the following main lemma.

\begin{lemma}
\label{lem:main}
Let $i \in \Good, (j,k) \in \Good_i$. Then, 
\begin{enumerate}
\item $S_j^{i-1} \cap S_k^{i-1} = \emptyset$ 
\item $ S_j^{i-1} \cup S_k^{i-1}
\subseteq S_j^i \cap S_k^i $.
\end{enumerate}
\end{lemma}
\begin{proof}
{\bf Part 1:} Let $ l \in S_j^{i-1}  \cap S_k^{i-1} $. Using standard
information theoretic relations it follows:
$$H(Y_k Y_j | Y_l \tx_i)  \leq  H(Y_k | Y_l \tx_i) +  H(Y_j  | Y_l
\tx_i)  \leq    2(a_i-1)\teps $$
Since $(j,k) \in \Good_i$ and from(\ref{eqn:main}),
$$  H(Y_k | \tx_i)  \geq  H(Y_k | X_i\tx_i) \geq  I(Y_k : Y_j |X_i\tx_i
) \geq 1 - \teps $$ 
Similarly $H(Y_j | \tx_i) \geq 1 - \teps$. 
Therefore again from(\ref{eqn:main}),
\begin{eqnarray*}
H(Y_jY_k | \tx_i) & = & H(Y_j | \tx_i) + H(Y_k | \tx_i) - I (Y_j
: Y_k | \tx_i)  \\
& \geq & 2 - 2\teps -  \teps = 2 -  3\teps  
\end{eqnarray*}

Now,
\begin{eqnarray*}
I( Y_l : Y_jY_k | \tx_i) & = &  H(Y_jY_k | \tx_i) - H(Y_jY_k | Y_l\tx_i) \\
& \geq &  2 - 3\teps - 2 ( a_i - 1) \teps  \geq 2 - 2(a_i + 1) \teps > 1 \mbox{ (since $a_i \leq \frac{1}{\tilde{2\epsilon}} - 2$) }
\end{eqnarray*}
This is a contradiction since $Y_l$ is a binary random variable. \\

{\bf Part 2:} We show $ S_j^{i-1} \cup S_k^{i-1} \subseteq S_j^i $ and $S_j^{i-1}
\cup S_k^{i-1} \subseteq S_k^i $ follows similarly. It is easily seen
that $S_j^{i-1} \subseteq S_j^i$. Let $l \in S_k^{i-1}$. Since $(j,k)
\in \Good_i$, from(\ref{eqn:main}),
$$H(Y_j | Y_k X_i \tx_i) = H(Y_j | X_i \tx_i) - I (Y_j : Y_k | X_i
\tx_i) \leq 1 - ( 1 - \teps) = \teps $$ 
Now,
\begin{eqnarray*}
H(Y_j | Y_l X_i \tx_i ) & \leq & H (Y_jY_k | Y_lX_i \tx_i )  \\
& = & H (Y_k | Y_l X_i \tx_i )  + H (Y_j | Y_lY_k X_i \tx_i )
\mbox{ (from (\ref{eq:chainent}))}\\ 
&\leq &  H (Y_k | Y_l \tx_i ) + H (Y_j | Y_k X_i\tx_i ) \\
& \leq &  (a_i-1) \teps +  \teps  = a_i \teps\\
\end{eqnarray*}
Hence $l \in S_j^i$ and therefore $S_k^{i-1} \subseteq S_j^i$.
\end{proof}

Our theorem now finally follows. \\ \\
\begin{proof} {\bf[Theorem~\ref{thm:main}]}
Let $i \in \Good$. Since $\epsilon \leq \frac{c^2}{8n^2}$,
Claim~\ref{claim:nice} holds. Lemma~\ref{lem:main} implies that for
$(j,k) \in \Good_i$, either $|S_j^i| = 2 |S_j^{i-1}|$ or $|S_k^i| = 2
|S_k^{i-1}|$.  Then, 
\begin{equation}
\label{eqn:Ssize}
\sum_j \log |S_j^{i -1}| + |\Good_i| \leq
\sum_j \log |S_j^i|
\end{equation}

Let $\tilde{i}$ be the largest $i \in
\Good$. Now,
\begin{eqnarray*}
(\frac{n}{40c} - 2 )\frac{m}{8c} & \leq & \sum_{i \in \Good}
\frac{|M_i|}{2} \mbox{ (from Lemma~\ref{lem:largematching})}\\ 
& \leq & \sum_{i \in \Good} |\Good_i| \mbox{ (from
Claim(\ref{claim:markov}) and $\delta_2 = 1/2$)}\\ 
& \leq & \sum_j \log |S_j^{\tilde{i}}| \mbox{
(from(\ref{eqn:Ssize}))}\\ 
& \leq & m \log m \\
\Rightarrow  m & \geq & 2^{(\frac{n}{40c} - 2 )\frac{1}{8c}} \geq  
2^{\frac{n}{320c^2} - 1}    
\end{eqnarray*}
\end{proof}

\section{Conclusion} We have attempted here a classical proof of an
already known theorem~\cite{ronald:ldc} which however has been shown
using quantum arguments. We hope that the arguments used here are
helpful in matching the result derived using quantum arguments. The
need for a classical proof is also due to the fact that the quantum
arguments do not help us to derive interesting bounds for codes with
higher number of probes, in particular even for $3$-probe codes.
\\ \\ 
{\bf \large Acknowledgment:} We thank Ashwin Nayak, Jaikumar Radhakrishnan,
Pranab Sen and Ronald de Wolf for useful discussions  and comments.

\end{document}